\documentclass{elsart}

\makeatletter
\newif\if@restonecol
\makeatother

\usepackage{graphicx,amssymb}
\usepackage[linesnumbered,ruled,vlined]{algorithm2e}
\usepackage{amsmath}
\usepackage[usenames]{color}
\usepackage{longtable}
\usepackage{booktabs}
\usepackage{url}
\usepackage{mathrsfs}
\usepackage{amsfonts}
\usepackage{marvosym}
\usepackage{lscape}
\usepackage{enumerate}
\usepackage{caption}
\usepackage{subcaption}
\usepackage{multirow}
\usepackage{verbatim}
\usepackage{pst-node}
\usepackage{lineno}
\newtheorem{theorem}{Proposition}
\SetKw{Continue}{continue}
\DeclareMathOperator*{\argmax}{\arg\!\max}
\begin{document}
\begin{frontmatter}
\title{Towards Effective Exact Algorithms for the Maximum Balanced Biclique Problem}
\author[label1]{Yi Zhou}, 
\author[label1]{Andr\'e Rossi}, 
\author[label1,label2]{Jin-Kao Hao\corauthref{cor}} %\ead{hao@info.univ-angers.fr}
\address[label1]{LERIA, University of Angers, 2 bd Lavoisier, 49045 Angers, France}
\address[label2]{Institut Universitaire de France, Paris, France}
%\cortext[cor1]{Corresponding author.}
\corauth[cor]{Corresponding author.} \ead{hao@info.univ-angers.fr}

\begin{abstract}
The Maximum Balanced Biclique Problem (MBBP) is a prominent model with numerous applications. Yet, the problem is NP-hard and thus computationally challenging. We propose novel ideas for designing effective exact algorithms for MBBP. Firstly, we introduce an Upper Bound Propagation procedure to pre-compute an upper bound involving each vertex. Then we extend an existing branch-and-bound algorithm by integrating the pre-computed upper bounds. We also present a set of new valid inequalities induced from the upper bounds to tighten an existing mathematical formulation for MBBP. Lastly, we investigate another exact algorithm scheme which enumerates a subset of balanced bicliques based on our upper bounds. Experiments show that compared to existing approaches, the proposed algorithms and formulations are more efficient in solving a set of random graphs and large real-life instances.

\end{abstract}
\end{frontmatter}

\section{Introduction}
\label{Introduction}

Given a bipartite graph $G=(U,V,E)$ with two disjoint vertex sets $U$, $V$ and an edge set $E\subseteq U\times V$, a biclique $A\cup B$ (or $(A,B)$) is the union of two subsets of vertices $A\subseteq U$, $B\subseteq V$ such that $\forall u\in A, \forall v\in B$, $\{u,v\}\in E$. In other words, the subgraph induced by vertex set $A \cup B$ is a complete bipartite graph. If $|A|=|B|$, then biclique $(A,B)$ is a balanced biclique. The Maximum Balanced Biclique Problem (MBBP) is to find a balanced biclique $(A,B)$ of maximum cardinality. As $|A|=|B|$ holds for a balanced biclique $(A,B)$, MBBP is then to find the maximum half-size balanced biclique. MBBP is a special case of the conventional maximum clique problem \cite{wu2015review}.

MBBP is a prominent model with a large range of applications, such as nanoelectronic system design \cite{al2007defect,tahoori2006application}, biclustering of gene expression data in computational biology \cite{cheng2000biclustering} and PLA-folding in the VLSI theory \cite{ravi1988complexity}. In terms of computational complexity, the decision version of MBBP is NP-Complete \cite{garey1979computers,alon1994algorithmic}, though the maximum biclique problem in bipartite graphs (without requiring $|A|=|B|$) is polynomially solvable by maximum matching algorithm \cite{cheng2000biclustering}. 

Considerable effort has been devoted to the pursuit of effective algorithms for MBBP, both theoretically and practically. Heuristic algorithms represent the most popular approach for MBBP, though they do not guarantee the optimality of the final solution found. The majority of existing heuristic algorithms solve the equivalent maximum balanced independent set (a vertex set such that no two vertices are adjacent) problem in the complement graph, rather than directly seeking the maximum balanced biclique from the given graph. For example, several greedy heuristic algorithms were proposed based on vertex-deletion on the complement graph from 2006 to 2014 \cite{al2007defect,tahoori2006application,yuan2011low,yuan2014fast}, while in \cite{yuan2015new}, an evolutionary algorithm combining structure mutation and repair-assisted restart was studied.
%A recent heuristic algorithm in \cite{zhou2016irs}, which employs graph reduction and local search, directly searches maximum balanced biclique from original graph. It performs quite well for large dense graphs and very large real-life networks.

On the other hand, according to our literature review, there are only two studies on exact algorithms in the literature. In \cite{tahoori2006application}, a recursive exact algorithm for searching a maximum balanced independent set with a given half-size in the complement graph was proposed. However, the computational time of this algorithm becomes prohibitive when the number of vertices of the given graph exceeds (32,32). In \cite{mccreesh2014exact}, a branch-and-bound (B\&B)  algorithm for MBBP for general graphs (including non-bipartite graphs) was studied. The algorithm incorporates a clique cover technique for upper bound estimation (an equivalent technique of using graph coloring to estimate the upper bound for the maximum clique problem) and  employs lex symmetry breaking techniques for general graphs. As far as we know, this algorithm is currently the best performing exact algorithm, even though the bounding technique and symmetry breaking techniques are only effective for non-bipartite graphs. 

In addition to specifically designed exact algorithms, the general Mixed Integer Programming (MIP) constitutes an interesting alternative for addressing hard combinatorial problems such as MBBP. Commercial MIP solvers, like IBM CPLEX, can even solve some hard instances which cannot be handled by other approaches. Meanwhile, the success of a MIP solver highly depends on the tightness of the mathematical formulation of the problem. For MBBP, a MIP formulation has been proposed in \cite{dawande2001bipartite}, it is based on the complement graph. Another mathematical formulation which defines the constraints on the original graph was presented in \cite{yuan2015new}. However, this formulation was not applicable for MIP solvers as it contains non-linear constraints. 

%\subsection{Contribution and Paper Organization}
In this work, we introduce new ideas for developing effective exact algorithms for MBBP, which can be applied to solve very large  MBBP instances from applications like social networks. Our main contributions can be summarized as follows.
\begin{itemize}
\item We elaborate an Upper Bound Propagation (UBP) procedure inspired from \cite{soto2011three}, which produces an upper bound of the maximum balanced biclique involving each vertex in the bipartite graph. UBP propagates the initial upper bound involving each vertex and achieves an even tighter upper bound for each vertex. UBP is independent from the search procedure and is performed before the start of the algorithm. An extended exact algorithm, denoted by (ExtBBClq), is proposed by taking advantage of UBP to improve BBClq, the branch-and-bound algorithm introduced in \cite{mccreesh2014exact}.

\item Based on the upper bounds returned by UBP, we introduce new valid inequalities to tighten the MIP formulation of MBBP introduced in \cite{dawande2001bipartite}. Our computational experiments suggest that using the tightened model improves the performance of the MIP solver CPLEX.

\item We also present a new exact algorithm (ExtUniBBClq) to supplement the family of B\&B based algorithms for MBBP. Unlike BBClq which goes through every possible balanced biclique, the new algorithm only enumerates the possible partial sets (half-sets) of the balanced bicliques in the graph. ExtUniBBClq also integrates UBP as a pre-processing procedure and performs generally well for the benchmark instances.
\end{itemize}

The reminder of the paper is organized as follows. Section \ref{sec_preliminaries} introduces the notations that will be used throughout the paper and Section \ref{sec_review_bbclq} reviews the BBClq algorithm introduced in \cite{mccreesh2014exact}. In Section \ref{sec_ubp}, we present our Upper Bound Propagation procedure for upper bound estimation and explain how to use it to improve BBClq. Then, in Section \ref{sec_math_form}, we show how the upper bounds can lead to new valid inequalities to tighten the MIP formulation of \cite{dawande2001bipartite}. Furthermore, we introduce the novel ExtUniBBClq algorithm in Section \ref{sec_unibbclq}. Computational results and experimental analyses are presented in Section \ref{sec_experiments}, followed by conclusions and future working directions.

\section{Notations}
\label{sec_preliminaries}
\begin{figure}[ht!]
	%\centering\scalebox{0.8}{\includegraphics[scale=0.7]{demo2}}	
	\begin{pspicture*}(-.5,-.5)(6.5,2.5)
	\pnode(0,2){1}\pscircle(0,2){.35}\rput(0,2){$1$}
	\pnode(1.5,2){2}\pscircle(1.5,2){.35}\rput(1.5,2){$2$}
	\pnode(3,2){3}\pscircle(3,2){.35}\rput(3,2){$3$}
	\pnode(4.5,2){4}\pscircle(4.5,2){.35}\rput(4.5,2){$4$}
	\pnode(6,2){5}\pscircle(6,2){.35}\rput(6,2){$5$}
	\pnode(0,0){6}\pscircle(0,0){.35}\rput(0,0){$6$}
	\pnode(1.5,0){7}\pscircle(1.5,0){.35}\rput(1.5,0){$7$}
	\pnode(3,0){8}\pscircle(3,0){.35}\rput(3,0){$8$}
	\pnode(4.5,0){9}\pscircle(4.5,0){.35}\rput(4.5,0){$9$}
	\pnode(6,0){10}\pscircle(6,0){.35}\rput(6,0){$10$}
	\ncline[nodesep=.35]{1}{6}
	\ncline[nodesep=.35]{1}{7}
	\ncline[nodesep=.35]{2}{7}
	\ncline[nodesep=.35]{2}{8}
	\ncline[nodesep=.35]{2}{9}
	\ncline[nodesep=.35]{3}{7}
	\ncline[nodesep=.35]{3}{8}
	\ncline[nodesep=.35]{3}{9}
	\ncline[nodesep=.35]{4}{6}
	\ncline[nodesep=.35]{4}{9}
	\ncline[nodesep=.35]{4}{10}
	\ncline[nodesep=.35]{5}{8}
	\ncline[nodesep=.35]{5}{10}
	\end{pspicture*}
	\centering\caption{A bipartite graph $G=(U,V,E)$, $U=\{1,2,3,4,5\}$, $V=\{6,7,8,9,10\}$.} \label{fig_demo}
\end{figure}
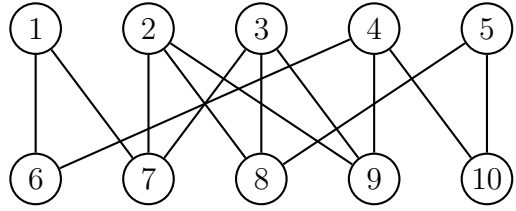

%Given a bipartie graph $G=(U,V,E)$ ($|U|\le|V|$ if not specifically stated), let $A$, $B$ such that $A\subseteq U$, $B\subseteq V$ and the induced graph $G[A\cup B]$ is complete (i.e., $E(G[A\cup B]) = A\times B$). A biclique $(A,B)$ is a balanced biclique if $|A|=|B|$, the half-size of a balanced biclique $(A,B)$ is the cardinality of $|A|$ (or $|B|$). For example, in Figure \ref{fig_demo}, $(\{2,3\},\{7,8\})$ is a balanced biclique of half-size of 2. Given a vertex $v$ in $G$, the set of vertices adjacent to $v$ is denoted by $N(v) = \{u|(v,u)\in E\}$, the degree of vertex $v$ ($|N(v)|$) is denoted by $deg_G(v)$. The upper bound involving vertex $v$, denoted by $ub_v$, is an upper bound of the half-size of the maximum balanced biclique containing vertex $v$. 

Given a bipartie graph $G=(U,V,E)$ ($|U|\le|V|$ if not specifically stated), let $(A,B) \subseteq (U,V)$ be a balanced biclique of $G$ (i.e., $|A|=|B|$). The half-size of the balanced biclique $(A,B)$ is the cardinality of $|A|$ (or $|B|$). For example, in Figure \ref{fig_demo}, $(\{2,3\},\{7,8\})$ is a balanced biclique of half-size of 2. For all $S \subseteq U \cup V$, $G[S]$ denotes the subgraph of $G$ induced by $S$. Given a vertex $v$ in $G$, the set of vertices adjacent to $v$ is denoted by $N(v) = \{u : \{v,u\}\in E\}$ and $deg_G(v) = |N(v)|$ is the degree of vertex $v$. The upper bound involving vertex $v$, denoted by $ub_v$, is an upper bound of the half-size of the maximum balanced biclique containing vertex $v$. For example, in Figure \ref{fig_demo}, a possible value for $ub_1$ could be 2, since $deg_G(1)=2$.

\section{Review of the BBClq algorithm}
\label{sec_review_bbclq}

Algorithm \ref{proc_bbexpand} shows the BBClq algorithm, which is a recursive exact algorithm introduced in \cite{mccreesh2014exact}. BBClq is adapted from a well-known B\&B algorithm for the maximum clique problem \cite{carraghan1990exact} and recursively builds up two sets $A$ and $B$ such that $(A,B)$ forms a biclique. The algorithm maintains a candidate set $C_A$ ($C_B$) that includes vertices which are eligible to move into $A$ ($B$) while ensuring that $(A,B)$ is a biclique (i.e., $C_A=\bigcap_{i\in B}N(i)$, $C_B=\bigcap_{i\in A}N(i)$). Initially, the algorithm sets $lb$, the global lower bound on the maximum biclique half-size to 0 and starts the search by calling BBClq$(G,\emptyset, \emptyset, U, V)$.

At each recursive call to BBClq, a vertex $v$ (called branch vertex) is moved from $C_A$ (lines 7-8). The algorithm then considers the branches (possibilities) of $v\in A$ in lines 9-12 and $v\notin A$ in the next {\tt while} loop. The bounding procedure (line 9) prunes the branch of $v\in A$ if the upper bound after estimation in this context is not larger than the global lower bound. The upper bound estimating method, which is classically a key point concerning the performance of a B\&B algorithm, will be introduced in the following section. If the current branch is not pruned, the search goes on by reconstructing $A'$ with a new vertex $v$ and $C'_B$ by filtering from $C_B$ those vertices not adjacent to $v$ (every vertex in $B$ must be adjacent to every vertex in $A$). After updating the two sets, the algorithm recursively calls BBClq in line 12, swapping the roles of $A$ and $B$, as $A$ and $B$ are extended alternatively for the sake of satisfying the balance requirement. The above process is repeated in the next recursive call of BBClq.

When the algorithm loops back to line 4, as we just mentioned, it explores another branch implying $v\notin A$. The {\tt while} loop stops when $C_A$ becomes empty or when the remaining vertices in $C_A$ do not allow to build a solution better than the global lower bound (lines 5-6). Besides, since $|A|+1=|B|$ or $|A|=|B|$ holds each time BBClq is called, we update the lower bound in lines 1-3 once $|A|>lb$ and store the incumbent solution $(A,B)$ as the best solution found so far. As a result, the best solution $(A^*,B^*)$ is an optimal biclique with $|A^*|=lb$ ($A^*\subseteq U$ or $A^*\subseteq V$), but it may not be totally balanced ($|A^*|-|B^*|\le 1$). Thus, in line 13, the procedure of retrieving the maximum balanced biclique (of half-size $lb$) from a biclique is accomplished by \textit{make\_balance()}. This procedure simply removes vertices from the larger set $A^*$ or $B^*$ until a balanced biclique is obtained.
 
Figure \ref{fig_demo} is now used to illustrate BBClq. Initially, $lb=0$ and BBClq$(G, \emptyset, \emptyset, U, V)$ is called. According to the minimal degree heuristic in \cite{mccreesh2014exact}, vertex $1$ is chosen as the first branch vertex. Clearly, the current upper bound is greater than 0, the algorithm proceeds to BBClq$(G, \emptyset, \{1\}, \{6,7\}, U\setminus \{1\})$ to explore the solutions containing vertex $1$. As a result, the solution $(\{1\},\{6\})$ is found and $lb$ is updated to 1. Likewise, the algorithm selects $5$ as the second branch vertex in the following loop, proceeds to BBClq$(G, \emptyset, \{5\}, \{8,10\}, U\setminus\{1,5\})$ if no upper bounding technique is applied. We can see that this recursive call to BBClq has to explore the case of expanding the given biclique by adding vertex 8 or 10. However, with the upper bounding estimating technique proposed in this paper, the call of BBClq$(G, \emptyset, \{5\}, \{8,10\}, U\setminus\{1,5\})$ will not even start since the upper bound involving vertex 5 is 1 ($upper\_bound(\{5\})=lb=1$). The algorithm finds the optimal solution $(\{1,2\},\{4,5\})$ after the third loop (which explores $A=\{2\}$ and calls BBClq$(G, \emptyset, \{2\}, \{7,8,9\}, U\setminus\{1,5,2\})$). There will be no additional iteration as $|A|+|C_A|\le 2$ ($A=\emptyset, C_A=\{3,4\}$).

\begin{algorithm}[ht!]
	\caption{BBClq($G$, $A$, $B$, $C_A$, $C_B$), the B\&B algorithm for MBBP taken from \cite{mccreesh2014exact}}
	\label{proc_bbexpand}
	\KwIn{Graph instance $G=(U,V,E)$, $A$, $B$ - current sets that form a biclique, $C_A$, $C_B$ - the sets of eligible vertices that can be added to $A$ and $B$ respectively}
	\KwOut{A maximum balanced biclique of $G$.}
	\If{$|A| > lb$}{
		$lb \gets |A|$ \;
		Record current best biclique in $(A^*,B^*)$\;			
	}
	\While{$C_A \neq \emptyset$}{
		\If{$|A| + |C_A| \leq lb$}{
			\Return
		}
		$v \gets branch\_vertex(C_A) $\;
		$C_A \gets C_A\setminus\{v\}$\;
		\If{$upper\_bound(A\cup\{v\}) > lb$}{
			$A'\gets A\cup\{v\}$\;
			$C'_B\gets C_B\cap N(v)$\;
			BBClq($G$,$B$, $A'$, $C'_B$, $C_A$)
		}
	}
	\Return {make\_balance$(A^*, B^*)$}
\end{algorithm}

\section{Upper bound propagation and its use to improve BBClq}
\label{sec_ubp}

We introduce in this section our Upper Bound Propagation procedure (UBP) which is then used as a pre-processing technique to reinforce the BBClq algorithm presented in the last section.

\subsection{The upper bound propagation procedure}
\label{UBP}

The original BBClq algorithm calculates a clique cover (based on addressing the graph coloring problem on the complement graph) to estimate the upper bound in a general graph relying on the fact that sets $A$ and $B$ are independent sets. However, when the given graph is bipartite, the upper bound found by this technique is trivial as two vertex sets are initially independent sets. Here, we introduce our Upper Bound Propagation to produce, for each vertex, an upper bound on the half-size of any maximum balanced biclique involving that vertex. UBP is based on the following propositions.

\begin{theorem}
\label{prop_degree}
For each vertex $v\in U\cup V$, $deg_G(v)$ is an upper bound on the maximum half-size balanced biclique involving $v$. 
\end{theorem}

This proposition is obviously true since the half-size of a balanced biclique cannot exceed the degree of any vertex in the biclique.

\begin{theorem}
\label{prop_cns}
Given a vertex $v \in U$, let $w_{vu} =|N(v)\cap N(u)|, \forall u\in U$. Let $y_v$ be the maximum integer such that there exists at least $y_v$ vertices in $\{w_{vu},u\in U\}$ satisfying $w_{vu}\geq y_v$, then $y_v$ is an upper bound on the maximum half-size balanced biclique involving $v$.
\end{theorem}

\textit{Proof}: Clearly, in the maximum balanced biclique $(A,B)$ involving $v \in A$, for any vertex $u \in A$ (including $v$), we have $B\subseteq N(v)\cap N(u)$. Therefore, the maximum possible value $y_v$ such that $y_v$ vertices in $U$ share at least $y_v$ adjacent vertices with $v$ is an upper bound involving $v$. Note that this proposition also holds given any vertex in $V$.
	
\begin{theorem}
\label{prop_ub_propagate}
Given a vertex $v\in U\cup V$, let $z_v$ be the largest integer such that there exists $z_v$ vertices in $N(v)$  having upper bounds at least $z_v$. Then $z_v$ is an upper bound on the maximum half-size balanced biclique involving $v$.
\end{theorem}

\textit{Proof}: We prove this proposition by contradiction. Suppose $z_v$ is not an upper bound, then there exists a balanced biclique $(A',B')$ involving $v\in A'$ of half-size $z'_v$ such that $z'_v>z_v$, implying that all the $z'_v$ vertices in $B'$ ($B'\subseteq N(v)$) must have an upper bound of at least $z'_v$ (i.e., $\forall u \in B, ub_v \ge z'_v$), which contradicts the condition that $z_v$ is the maximum integer such that there exists in $N(v)$ at least $z_v$ vertices having $z_v\geq ub_v$. 

Consider the example of Figure \ref{fig_demo}, according to Proposition \ref{prop_degree}, we have $ub_1 = ub_5 = 2$, $ub_2 = ub_3 = ub_4 = 3$. Then, following Proposition \ref{prop_cns}, $ub_1$ can be improved (decreased) to $1$ since $w_{12}=w_{13}=w_{14} =1$, $w_{15}=0$ ($y_1=1$). Similarly $ub_2,ub_3,ub_4, ub_5$ can also be improved to 2, 2, 1, 1 respectively. By Proposition \ref{prop_ub_propagate}, it can be deduced that $ub_6=1$ and $ub_7=2$ ($z_6=1$, $z_7=2$), which are better upper bounds than the degrees.

Based on these proposition, we devise the UBP procedure (see Algorithm \ref{proc_ub_propagate}) to calculate an upper bound involving each vertex. Initially $ub_v$ is set to $deg_G(v)$, then the upper bound of each vertex in $U$ is improved according to Proposition \ref{prop_cns} (lines 2-9). From line 10 to the end of Algorithm \ref{proc_ub_propagate}, the procedure aims at propagating the upper bound based on Proposition \ref{prop_ub_propagate} until the upper bounds cannot be improved any more. The propagation procedure is guaranteed to converge as the upper bounds cannot be smaller than 0. Experiments in Section \ref{sec_experiments} show that, for both random and real-life large instances, UBP converges very fast, only in a limited number of iterations.

In both lines 7 and 14, we use binary search to find, for a given set $I$ of integers, the maximum element $x \in I$ such that there are at least $x$ integers in $I$ that are larger than or equal to $x$. The procedure works as follows: first, $I$ is sorted by decreasing order, then, an iteration starts by comparing the middle element with its index in $S$ (i.e., its position in the sorted list). If the middle element is greater (respectively lesser) than its index, the next iteration proceeds with the second half (respectively the first half) of $I$. This binary search procedure based on dichotomy performs at most $\log_2(|I|)$ operations.

Actually, we can also tighten the initial upper bound involving each vertex in $V$ by repeating the process in lines 2-9 after replacing $U$ with $V$ before the propagating procedure (lines 10-17) starts. However, this procedure requires considerable memory and time especially for large graphs. For example, the matrix representing $w_{ij}, (i,j)\in V\times V$ requires a memory of $O(|V|^2)$, the computational time of computing $y_v$ ($v\in V$) is bounded by $O(|U|\times(max_{v\in V}deg_G(v))^2)+|V|\times\log(|V|)$. Thus the overhead is not negligible. As a compromise, we set a threshold on the size of the vertex set. We apply the procedure of lines 2-9 to improve the upper bound involving each vertex only when the cardinality of the vertex set ($U$ or $V$) is less than the threshold. In the following experiments, the threshold has empirically been set to 30000.

\begin{algorithm}[h]
	\caption{Upper bound propagation procedure}
	\label{proc_ub_propagate}
	\KwIn{Graph instance $G=(U,V,E)$ }
	\KwOut{An upper bound vector $ub$ for each vertex in $G$.}

	$\forall{v} \in U\cup V, ub_v \gets deg_G(v)$\;
	$\forall (v,u) \in U\times U, w_{vu} \gets 0$\;
	\For{$k \in V$}{
		\For {$(v,u) \in N(k)\times N(k)$}{
		 	$w_{vu} \gets w_{vu}+1$\;
		 }
	}
	\For {$v \in U $}{
		Binary search for the largest integer $y_v$ such that $|\{u\in U: w_{vu} \geq y_v\}| \geq y_v$\;
		\If{$y_v < ub_v$}{
			$ub_v \gets y_v$\;
		}
	}
	$stable \gets false$\;
	\While{$stable \neq true$}{
		$stable \gets true$\;
		\For{$v \in U\cup V$}{
			Binary search for the largest integer $z_v$ such that $|\{u\in N(v): ub_u \geq z_v\}|\geq z_v $\;
			\If{$z_v < ub_v$}{
				$ub_v \gets z_v$\;
				$stable \gets false$\;
			}
		}
	} %end of while
	\Return $ub$
\end{algorithm}

To see how tight the upper bounds provided by UBP are, consider the example of Figure \ref{fig_demo}, the final upper bound achieved by UBP is $ub_v={1,2,2,1,1}$ for $v \in U$ and $ub_v={1,2,2,2,1}$ for $v\in V$. These upper bounds are actually all tight.

\subsection{Combining UBP with BBClq: ExtBBClq}
\label{sec_combine_ubp_bbclq}

As UBP is independent of the search algorithm, we use it as a pre-processing procedure for BBClq to obtain an extended version named ExtBBClq. In ExtBBClq, we use the same branching heuristic as in the original BBClq algorithm: the vertex of the minimum degree in $C_A$ is given the highest priority for branching. To efficiently implement ExtBBClq, we sort the arrays $N(v)$ ($\forall v\in U\cup V$) in ascending order of index number before the beginning of BBClq, so that the intersection operation in line 11 (Algorithm \ref{proc_bbexpand}) can be accomplished in $O(|C_B|*log(|N(v)|))$ asymptotic time by binary search. More importantly, to make use of the upper bound information calculated by UBP, in ExtBBClq, instead of calculating the upper bound by calling the upper bound estimation method (i.e., $upper\_bound(A\cup\{v\})$ in line 9, we use the pre-computed $ub_v$ returned by UBP as the upper bound in the current branch.

\section{A tighter mathematical formulation}
\label{sec_math_form} 

In this section, we propose a tightened mathematical formulation for MBBP that takes advantage of the UBP procedure. Let us first recall the mathematical formulation of MBBP introduced in \cite{dawande2001bipartite}:

\begin{equation}\label{f1}
\max \ \ \omega(G)=\sum_{i=1}^{|U|}{x_i}
\end{equation}
subject to:
\begin{equation}\label{c1}
x_i + x_j \leq  1, \forall\{i,j\}\in \bar{E}
\end{equation}
\begin{equation}\label{c2}
\sum_{i=1}^{|U|}{x_i} - \sum_{i=|U|+1}^{|U|+|V|}{x_i} = 0
\end{equation}
\begin{equation}\label{c3}
x_i \in \{0, 1\},  \forall i \in U \cup V
\end{equation}

where each vertex of $U \cup V$ is associated to a binary variable $x_i$ indicating whether the vertex is part of the biclique, $\bar{E}$ is the set of edges in the complement bipartite graph of $G$. Constraint (\ref{c1}) requires that each pair of non-adjacent vertices cannot be selected at the same time (i.e., the solution must form a biclique). Constraint (\ref{c2}) enforces that the biclique is balanced.

To make use of the upper bounds returned by UBP, let $S^\ell \subseteq U$ (or $S^\ell \subseteq V$) be the set of all the vertices in $U$ (respectively in $V$) such that $ub_i \le \ell$ ($\ell$ is a positive integer) for all $i \in S^\ell$. Then the following inequality is valid:

\[
\displaystyle\sum_{i \in S^\ell} x_i \le \ell
\]

Indeed, the vertices in $S^\ell$ can only be involved in balanced cliques having half-size less than $\ell$. We consider this inequality for $\ell = \displaystyle\max_{i \in U} ub_i$ as it dominates the inequalities associated with lower values of $\ell$.

Before tightening this inequality, we observe that since $\ell = \displaystyle\max_{i \in U} ub_i$, we have $S^\ell = U$. Then for each $u \in U$ such that $ub_u < \ell$, or equivalently for all $u \in S^{\ell-1}$, we can lift the term associated with $u$:

\[
(\ell - ub_u + 1) x_u + \displaystyle\sum_{i \in U\backslash\{u\}} x_i \le \ell \hspace {3ex}\forall u \in S^{\ell-1}
\]

Let $T^{\ell-1}_u$ be any maximal subset of $S^{\ell - 1}$ containing $u$ such that for all $i$ and $j \in T^{\ell-1}_u$, then $N(i) \cap N(j)$ is empty. The term `maximal subset' means that no vertex $i \in S^{\ell-1}$ can be added to $T^{\ell-1}_u$.

We can deduce the following valid inequality:

\[
\displaystyle\sum_{i \in T^{\ell-1}_u} (\ell - ub_i + 1) x_i + \displaystyle\sum_{i \in U\backslash T^{\ell-1}_u} x_i \le \ell \hspace {3ex}\forall u \in S^{\ell-1}
\]

It can be observed that the valid inequalities built from two vertices $u$ and $v$ of $S^{\ell-1}$ may possibly be identical, especially if $v \in T^{\ell-1}_u$. This is not an issue since modern solvers remove duplicate constraints automatically during presolving.

Naturally, the lower $ub_i$ is, the tighter these inequalities are. Consider the example of Figure \ref{fig_demo}, since the upper bound involving each vertex is given by UBP, we can produce the following valid inequalities ($\ell = 2$):
\begin{itemize}
\item Vertex 1 (and also 4) leads to $2x_1 + x_2 + x_3 + 2x_4 + x_5 \le 2$
\item Vertex 5 leads to $2x_1 + x_2 + x_3 + x_4 + 2x_5 \le 2$
\item Vertex 6 leads to $2x_6 + x_7 + x_8 + x_9 + x_{10} \le 2$
\item Vertex 10 leads to $x_6 + x_7 + x_8 + x_9 + 2x_{10} \le 2$
\end{itemize}

The LP relaxation of the original formulation (1)-(4) yields an objective of 2.5, and nearly all the variables are fractional. Adding these four inequalities yields an objective of 2 and an integer solution, which proves to be optimal.

%It is suggested that these inequalities could be faces (or facets) and may be useful in solving larger instances of the problem.

\section{A novel MBBP algorithm ExtUniBBClq}
\label{sec_unibbclq}

We observe that for any biclique $(A, B)$ such that $A\subseteq U, B \subseteq V$ and $|A|\leq|B|$, the maximum balanced biclique in subgraph $G[A\cup B]$ is $(A,B')$ with $B'\subseteq B$, and the maximum half-size of any $(A,B')$ is still $|A|$. In other words, the  half-size of the maximum balanced biclique in $G=(U,V,E)$ is the cardinality of maximum subset $A\subseteq U$ which satisfies $|\bigcap_{i\in A}{N(i)}| \geq |A|$. As a result, instead of building the two sets of balanced biclique alternatively, we can directly enumerate the eligible subset $A$ from $U$ (or $B$ from $V$) such that $|\bigcap_{i\in A}{N(i)}| \geq |A|$. Based on this observation, we propose a new algorithm (Algorithm \ref{proc_unibbexpand}) which builds the maximum eligible subset from $U$ (as $|U|\leq |V|$).

\begin{algorithm}[h]
	\caption{ExtUniBBClq($G$, $A$, $C_A$, $B$), a new B\&B procedure for MBBP based on enumerating one vertex set.}
	\label{proc_unibbexpand}
	\KwIn{Graph instance $G=(U,V,E)$ , $A$ - the current subset of $U$, $C_A$ - the candidate subset of $U$, $B$ - the common neighbors of vertices in $C$ ,i.e., $B=\bigcap_{i\in C}N(i)$ }
	\KwOut{A maximum balanced biclique of $G$}
	\If{$|A|\leq |B|$ AND $|A| > lb $}{
		$lb \gets |A|$ \;
		Record current best biclique $(A^*, B^*)$\;
	}
	\If{$|A|\geq |B|$ OR $|B| \leq lb$}{
		\Return{make\_balance$(A^*, B^*)$}
	}
	\While{$C_A \neq \emptyset$}{
		\If{$|A| + |C_A| \leq lb$}{
			\Return
		}	
		$v \gets \argmax_{i\in C_A}ub_i$\;
		$C_A \gets C_A\setminus \{v\}$\;
		\If{$ub_v \leq lb$}{
			\Return
		}			
		$A' \gets A\cup\{v\}$\;
		$B' \gets B \cap \{u\in N(v): ub_u>lb \}$\;
		 ExtUniBBlq($G$, $A'$, $C_A$, $B'$)
	}
	\Return {make\_balance$(A^*, B^*)$}
\end{algorithm}

%\begin{algorithm}[h]
%\caption{UniBBClq, an algorithm for MBBP with pre-computed upper bounds.}
%\label{algorithm_UniBBclq}
%\KwIn{Graph instance $G(U,V,E)$}
%\KwOut{The half size of maximum balanced biclique}
%\Begin{
%$lb\gets 0$  \tcc*[r]{global lower bound}
%Pre-compute upper bound $ub_v$  for all $v \in U$ \; 
%Re-sort $U$ by non-increasing order of $ub_v$\;
%UniBBexpand($G, \emptyset, U, V$)
%} %end of algorithm
%\Return{$\omega$}
%\end{algorithm}

%UniBBClq starts with an upper bound estimation procedure rather than directly call the recursive search procedure UniBBexpand (Procedure \ref{proc_unibbexpand}). In line 3, the upper bounds of the maximum half-size of balanced biclique involving vertex $i$ is pre-computed and reserved in corresponding $ub_i$ (the details are elaborated in the following section). The algorithm reorders the vertex of $U$ in non-increasing order by its associate upper bound. As UniBBexpand gives preference in branching the vertex with smaller index, it  branches the vertex associate with largest upper bound.
 
The framework of ExtUniBBClq is similar to Algorithm \ref{proc_bbexpand} except that ExtUniBBClq only builds the set $A$ recursively such that $(A,B)$ forms a clique and $|A|\leq|B|$. Moreover, in ExtUniBBClq, we make use of the upper bound involving each vertex returned by UBP. Therefore, UBP has to be called before the start of ExtUniBBClq. For each call of ExtUniBBClq, a current set $A$, as well as the candidate set $C_A$ that contains vertices that can be moved into $A$, and set $B$ which is the common adjacent vertices of vertex in $A$ (i.e., $B=\bigcap_{i\in A}N(i)$) are given. The algorithm initializes $lb$ to 0 and begins from ExtUniBBClq$(G, \emptyset, U, V)$.

As in BBClq, in each call of ExtUniBBClq, a branch vertex $v$ (with the maximum upper bound) is moved out from $C_A$ (lines 9-10) and the algorithm goes to two branches: the branch where $v\in A$ before the end of current loop (lines 11-15) and another branch where $v\notin A$ in the next loop. In lines 11-12, when the upper bound associated with vertex $v$ is not larger than the lower bound, ExtUniBBClq stops the current search immediately once $ub_i$ is the largest upper bound of all vertices in $C_A$. In lines 13-15, the search goes on by expanding $A'$ and rebuilding $B'$. Note that we filter out unpromising vertices from $N(v)$ which have an upper bound not larger than the lower bound. In the end of the loop, ExtUniBBClq is called to further enlarge set $A$. After it returns, the algorithm moves to the next loop, entering another branch with $v\notin A$. In lines 7-8, the search is stopped when $C_A$ is not large enough to build a better solution. In lines 5 and 13, the \textit{make\_balance} procedure is called so that the final solution is a strictly balanced biclique of half-size $lb$.

At the start of each call to ExtUniBBClq, we update the lower bound if it is needed (lines 1-3) and terminate the current search if $A$ is not eligible ($|A|\le |B|$) or $|B|$ is not larger than the lower bound. For an efficient implementation, we pre-sort the array which represents $U$ (the initial $C_A$) in ascending order of the upper bound involving each vertex. Consequently, the last element in the array will always be the vertex with the largest upper bound. We also sort the arrays representing $V$ (the initial $B$) and $N(v)$ ($\forall v \in U$) in ascending order of the index of each vertex so that the intersection operation in line 14 can be accomplished in linear time.

We illustrate the principle of this procedure by using the example of Figure \ref{fig_demo} again. Firstly, the lower bound $lb$ is initialized to 0 and ExtUniBBClq$(G,\emptyset, U, V)$ is then called. In the first {\tt while} loop, vertex $2$ is selected as the branch vertex as $ub_2=2$ is the largest upper bound of all vertices in $U$; then we get $A=\{2\}$, $C'_A =\{1,4,5,3\}$, $B'= \{7,8,9\}$. The next call to ExtUniBBClq expands the incumbent set $A=\{2\}$, which leads to a solution $(\{2,3\},\{7,8\})$ (and $lb$ is updated to 2). The second {\tt while} loop which branches on vertex $3$ and builds candidate set $C'_A=\{1,4,5\}$, is stopped earlier as the largest upper bound ($ub_3=2$) is equal to $lb$. Thus, the whole search stops, returning $(\{2,3\},\{7,8\})$ as the optimal solution. 

\section{Computational experiments}
\label{sec_experiments}

This section is dedicated to a computational evaluation of the proposed algorithms for MBBP, based on the following two sets of benchmark graphs which are commonly used in the literature \cite{hardiman2013estimating,yuan2015new}.

\begin{itemize}
\item[-] \textbf{Random graphs}. In these graphs, every possible edge occurs independently with fixed probability. For each graph, there are $n$ vertices in each vertex set (i.e., $n=|U|=|V|$) and the probability that an edge exists between a pair of vertices $(u,v)\in U\times V$ is $p$ ($p \in [0,1]$). A theoretical analysis in \cite{dawande2001bipartite} showed that the maximum half-size of the balanced bicliques in such graphs is in the range $[\frac{\ln n}{\ln(1/p)},\frac{2 \ln n}{\ln(1/p)}]$ with high probability (when $n$ is sufficiently large).
\item[-] \textbf{Real-life networks}. This set includes 30 bipartite networks from the Koblenz Network Collection (KONECT) \cite{kunegis2013konect}, which contains hundreds of networks derived from real-life applications, including social networks, hyperlink networks, authorship networks, physical networks, interaction networks and communication networks. We summarize the main features of the selected instances in Table \ref{tbl_konect_info}, including the number vertices and edges. Irrelevant information for solving MBBP, such as multiple edges, vertex or edges weights have been filtered out. 
\end{itemize}

	\begin{table}
	%\renewcommand\arraystretch{0.5}
	%\linespread{0.8} 
	\begin{scriptsize}
	\caption{The basic information of selected real-life networks.	\label{tbl_konect_info}}	
	\begin{tabular}{p{2.5cm}|ccc|p{2.5cm}|ccc}
	\hline
	instance& $|U|$ &$|V|$ &$|E|$ & 	instance& $|U|$ &$|V|$ &$|E|$ \\
	\hline
actor-movie	&127823 & 383640	 &1470418 &edit-dewiki			&425842	&3195148 &57323775\\
bibsonomy-2ui  &	5794 & 767447&2555080		&edit-frwiktionary	&5017 &1907247 &7399298  \\
bookcrossing\_full-rating	& 40523 &105278  &1149739 &escorts				&6624 &10106  &50632		\\
dblp-author	&1425813 &4000150	&8649016 &flickr-groupmemberships  &103631 &395979 &8545307 \\
dbpedia-genre  	&7783 &258934 &463497	&github  	&56519 &120867&440237\\
dbpedia-location		&53407	&172091	&293697 &gottron-trec &556077 &1173225  &83629405	\\
dbpedia-occupation	&101730	&127577	&250945 &jester1  		& 100 &73421&4136360		\\
dbpedia-producer  &48833 &138844 &207268		&moreno\_crime 		&551 &829 &1476	\\
dbpedia-recordlabel  &18421 &168337 &233286	&opsahl-ucforum  		&522 &899 &33720\\
dbpedia-starring  	&76099 &81085 &281396 &pics\_ut  	&17122 &82035 &2298816	\\
dbpedia-team			&34461	&901166	&1366466 &reuters	 &283911	 & 781265 &60569726\\
dbpedia-writer		&46213	&89356	&144340	&stackexchange-stackoverflow &96680 &545196  &1301942			\\
discogs\_affiliation	&270771	&1754823	&14414659	&unicodelang  	&254 &614 &1255			\\
discogs\_lgenre 		&6624 &10106 &50632		&wiki-en-cat&182947	&1853493 &3795796\\
discogs\_style  			&383 &1617943 &24085580 &youtube-groupmemberships  &30087 &94238 &293360			\\
	\hline
	\end{tabular}
	\end{scriptsize}
	\end{table}	
	
We compare the performance of 5 algorithms including both the existing approaches and the new approaches proposed in this work. The first 3 algorithms are B\&B algorithms.
\begin{itemize}
\item \textbf{BBClq}: the algorithm introduced in \cite{mccreesh2014exact}. However, compared with the original algorithm, symmetry breaking and clique cover techniques are removed as they are irrelevant for bipartite graphs. %The bit-parallel technique is also not applied based on the following considerations: first, the efficiency of bit-parallel is machine dependent; second, in a large graph, the operation of intersecting two large bit-vectors becomes less efficient and the adjacency matrix representation is also inevitable for bit-parallel.
\item \textbf{ExtBBClq}: the extended version of BBClq combining UBP with our new branching heuristic presented in Section \ref{sec_combine_ubp_bbclq}
\item \textbf{ExtUniBBClq}: the new algorithm introduced in Section \ref{sec_unibbclq}.
\end{itemize}

To compare the original mathematical formulation and the tightened formulation presented in this work, we use IBM CPLEX 12.6.1 to solve the benchmark instances with both formulations. 

\begin{itemize}
\item \textbf{Original}: the original mathematical formulation of MBBP from \cite{dawande2001bipartite}. 
\item \textbf{Tightened}: the formulation with the additional inequalities introduced in Section \ref{sec_math_form}. 
\end{itemize}

	\begin{table}[htbp]
	\centering
	%\renewcommand\arraystretch{0.5}
	%\linespread{0.8} 
	\begin{scriptsize}	
	\caption{Computational results of the 5 algorithms for the random graphs.\label{tbl_random_result}}		
	\begin{tabular}{cc|cc|ccc|cc}
	%\begin{tabular}{p{2.0cm}|p{0.4cm}|p{0.5cm}p{0.65cm}p{0.65cm}|p{0.8cm}p{0.8cm}}
	\hline
	\multirow{2}{*}{$n$} & \multirow{2}{*}{$p$} &\multicolumn{2}{c|}{UBP} & \multicolumn{3}{c|}{B\&B Algorithms} & \multicolumn{2}{c}{MIP}\\
	\cline{3-9}
	& & time & iter &BBClq  &ExtBBClq&ExtUniBBClq & Original & Tightened\\	
	\hline
50  &0.1  & 0.00 &3.1 &0.00		&0.00		&0.00		&4.02	&\textbf{0.41}  \\
50 &0.3  & 0.00 &2.8 &0.02		&\textbf{0.01}	&\textbf{0.01}	&4.62	&\textbf{4.70}  \\
50 & 0.5 & 0.00 &3.3  &0.45		&\textbf{0.15}	&0.3 	&\textbf{6.48}	&11.52  \\
50 & 0.7 & 0.00 &3.0   &24.19	&\textbf{5.12}	&11.60	&8.49  	&\textbf{7.47} \\
50 & 0.9 & 0.00 &3.2  &10174.28(5)	&\textbf{680.11} 	&4405.93(28) 	&0.39 	&\textbf{0.33}\\
100  & 0.1 & 0.00 &3.8  &0.01	&\textbf{0.00}	&\textbf{0.00}		&30.26	&\textbf{4.77}\\
100 & 0.3  & 0.00 &3.2  &0.96	&\textbf{0.42} 	&0.78	&\textbf{457.77} 	&510.60\\
100 & 0.5  & 0.01 &3.3  &118.97	&\textbf{32.22} 	&52.75	&7137.27		&\textbf{4392.06} \\
100 & 0.7  & 0.01 &3.1 	&[13.50-]	&\textbf{9540.81}(17)	&[13.73-]	&[13.57-20.30] 	&[13.40-20.00]  \\
100 & 0.9 & 0.00 &2.8   &[26.07-]	&[25.37-]	&[26.87-]	&\textbf{9977.23}(6)	&10358.21(4)  \\
150 & 0.1  & 0.00 &3.7  &0.06	&\textbf{0.04} 	&0.05	&305.04	&\textbf{225.67}  \\
150 & 0.3  & 0.01 &3.0 &11.28  	&\textbf{3.44}	&8.75	&9953.88(15)	&\textbf{9937.61}(14)   \\
150 & 0.5  & 0.02 &3.3 &4716.49 	&\textbf{933.95} 	&2281.69 	&[8.86-28.38]	&[8.89-28.46]  \\
150 & 0.7 & 0.04 &3.4 &[14.73-]	&[14.73-] 	&[15.00-] 	&[14.82-41.86]  	&[14.62-42.08]\\
150 & 0.9 & 0.05 &3.3  &[29.53-]	&[27.93-] 	&[30.23-]	&[35.04-55.58] 	&[34.71-55.54] \\
200 & 0.1 & 0.01 &3.3   &0.19	&\textbf{0.07}	&0.26 	&1725.24 	&\textbf{1239.80} \\
200 & 0.3  & 0.03 &3.2  &84.08	&\textbf{21.95}	&59.22	&[6.00-38.44]	&[6.00-22.52] \\
200 & 0.5   &0.05 &3.3 	&[9.97-]  	&\textbf{10761.34}(3)	&[10.03-]	&[8.95-55.68]	&[9.05-50.76] \\
200 & 0.7   &0.08 &3.3 &[15.30-]	&[15.23-]	&[15.93-]	&[15.45-64.32]	&[15.67-65.78] \\
200 & 0.9  &0.11 &3.3  &[31.93-]	&[30.10-]	&[32.60-]	&[38.70-79.19]	&[38.21-79.52] \\
%rand\_250\_0.1  &3* &0.34	&0.19	&0.15	&5996.09	&3722.35  \\
%rand\_250\_0.3  &7* &112.51	&69.17	&375.31	&-		&-  \\
%rand\_250\_0.5  &- &-(10)	&-(11)	&-(10)	&-		&-(10)  \\
%rand\_250\_0.7  &- &-(15)	&-(16)	&-(16)	&-(17)	&-  \\
%rand\_250\_0.9  &- &-(32)	&-(34)	&-(18)	&-(41)	&-(40)  \\
\hline
\end{tabular}
\end{scriptsize}
\end{table}

\begin{table}[htbp]
	%\renewcommand\arraystretch{0.5}
	%\linespread{0.8} 
	\begin{scriptsize}	
	\caption{Computational results of the 5 algorithms for KONECT instances.\label{tbl_konect_result}}		
	\begin{tabular}{c|c|cc|ccc|cc}
	%\begin{tabular}{p{2.0cm}|p{0.4cm}|p{0.5cm}p{0.65cm}p{0.65cm}|p{0.8cm}p{0.8cm}}
	\hline
	\multirow{2}{*}{instance} & \multirow{2}{*}{BEST} & \multicolumn{2}{c|}{UBP} & \multicolumn{3}{c|}{B\&B Algorithms} & \multicolumn{2}{c}{MIP}\\
	\cline{3-9}
	& &time & iter &BBClq &ExtBBClq&ExtUniBBClq & Original & Tightened\\	
	\hline
actor-movie						&8*  &5.54 &27 &6533.01	&1671.29 	&807.25	&-		&-\\
bibsonomy-2ui  					&8*  &1.56 &7  &491.36	 &13.84		&\textbf{9.13}	&- 		&-\\
bookcrossing\_full-rating		&13* &5.11 &33  &3102.66	&\textbf{426.37} &[10-]    &- &- \\
%brunson\_club-membershipp  		&3* &0.0		&0.0		&0.0		&9.47	&0.41	\\
%brunson\_corporate-leadership	&3*	&0.0		&0.0		&0.0		&10.18	&2.42	\\
%brunson\_revolution 			&3* &0.0 	&0.0		&0.0 	&-		&12.02	\\
%brunson\_south-africa 			&2* &0.0		&0.0		&0.0		&0.14 	&0.20	\\
%brunson\_southern-women  		&2* &0.0 	&0.0		&0.0		&0.23	&0.03	\\
dblp-author				&10* 	&19.86 &21  &[1-] 	&403.16	&\textbf{30.06}	&-	&-  \\
dbpedia-genre  			&7* 		&3.94 &9  &171.86	&\textbf{5.83}	&16.35	&-	&- \\ 
dbpedia-location  		&5* 		&0.18 &8  &633.98  &0.52 &\textbf{0.39}   &-  &-  \\
dbpedia-occupation  		&6* 		&0.27 &8  &909.03  &\textbf{1.29} &1.57    &-  & -\\
dbpedia-producer  		&6*  	&0.27 &11  &535.44	&\textbf{0.62}	&0.65	&-		&-\\
dbpedia-recordlabel  	&6* 		&24.33 &7  &214.45	&24.67	&\textbf{24.04}	&- 		&-	\\
dbpedia-starring  		&6* 		&1.07 &31  &530.61	&4.67	&\textbf{1.39} 	&-		&-\\
dbpedia-team  			&6* 		&3.08 &15  &2982.24 &\textbf{241.06}	&1170.25	 &-   &-		\\
dbpedia-writer  			&6* 		&0.19 &12  &283.16	&0.35 	&\textbf{0.23}	 &-   &-		\\
discogs\_affiliation		&26* 	&12.01 &17  &[1-]	&\textbf{1688.95} &[18-]  &-   &-    \\
discogs\_lgenre 			&15*		&0.06 &1  &37.08	&1.01	&\textbf{0.17}	&- 		&-	\\
discogs\_style  			&38 		&17.42 &22  &[23-] 	&[38-]	&[23-]	&- 		&-	\\
edit-dewiki  			&40   	&93.68 &23  &[1-]	&[40-] & [14-]   &- 		&-	\\
%edit-frwiki						&15	&-(1)	&-(15)	&-(15)	&-	&-	\\
edit-frwiktionary  		&19*		&9.56 &9  &944.21	&\textbf{152.5} &[19-]	&- 		&-	\\
escorts  				&6* 		&5.69 &6  &\textbf{7.68}	&10.05	&10.55	&- 		&-		\\
flickr-groupmemberships  &36 	&47.37 &36  &[34-]	& [36-]	&[18-]	&-		&-\\
github  					&12* 	&1.01 &16  &677.72 &\textbf{150.66}&[12-]	&-		&-\\
%gottron-reuters  				&25* &\textbf{1841.42}	&-(25)	&-(15)	&- 		&-	\\
gottron-trec				&83	 	&549.21 &35  &[33-]		&[38-]	&[83-]	&-		&-\\
jester1  				&100*	&1.86 &1  &1204.64	&1123.66	&\textbf{4.24}	&248.87&-\\
%lastfm\_song  					&27 &-(27)	&-(22)	&-(20)	&-		&-	\\
moreno\_crime 			&2* 		&0.05 &3  &\textbf{0.05}	&0.06	&0.06		&3483.58	 &\textbf{55.22}\\
%movielens-10m\_ut  				&9* &\textbf{6.4}		&21.24	&24.95	&-		&-	\\
%opsahl-collaboration  			&8* &\textbf{30.6}	&52.83	&52.6	&-		&-	\\
%opsahl-southernwomen  			&4* &0.0		&0.0		&0.0		&9.82	&7.27\\
opsahl-ucforum  			&5* 		&0.09 &10  &0.18	&\textbf{0.13}	&0.26	&[4-285] 	&[5-7] \\
pics\_ut  				&27 		&21.84 &7  &[27-]	&[23-]	&[23-] &-		&-	\\
reuters  				&39   	&611.76 &61  &[35-]	 &[39-]	&[12-]	&-		&-	\\
stackexchange-stackoverflow  &9* &4.62 &29  &4107.56	&\textbf{265.8}	&3690.8 &-		&-\\
unicodelang  			&4* 		&0.02 &5  &\textbf{0.01}	&0.02	&0.03 &1218.46 &\textbf{19.58}\\
wiki-en-cat				&14*		&7.67 &20	&[1-]	&\textbf{28.72}		&121.99	&-		&-\\
youtube-groupmemberships  &12*	&0.96 &21  &222.88	&\textbf{11.49}	&1784.76		&-		&-\\
	\hline
	\end{tabular}
	\end{scriptsize}
	\end{table}

All the experiments are conducted on a computer with an Intel Xeon\textsuperscript{\copyright} E5-2670 processor (2.5GHz and 2GB RAM) running CentOS 6.5. The BBClq, ExtBBClq and ExtUniBBClq algorithms are implemented in C++ and compiled with g++ using optimization option {\tt -O3}\footnote{The code of our algorithms will be available online.}. For each instance, a cut-off time limit of 3 hours (10800 seconds) is given to each trial.  When solving the DIMACS machine benchmark procedure ‘dfmax.c’\footnote{\url{dfmax: ftp://dimacs.rutgers.edu/pub/dsj/clique/}} without compilation optimization flag, the run time on our machine is 0.46, 2.68 and 10.70 seconds for graphs r300.5, r400.5 and r500.5 respectively.

The experimental results for random graphs are summarized in Table \ref{tbl_random_result}. For each configuration which is a pairwise combination of $n\in \{50,100,150,200\}$ (the cardinality of $U$ or $|V|$) and $p\in \{0.1,0.3,0.5,0.7,0.9\}$ (the edge density), we generated 30 graphs independently. Column ``time" reports the pre-processing time (the time of running UBP) and column ``iter" shows the average number of {\tt while} loops (i.e., lines 11-17, Algorithm \ref{proc_ub_propagate}) needed to stabilize the upper bound of all the vertices. For each algorithm, we report the average time consumed to solve the corresponding instance (the time for pre-processing is also included). Note that 0.00 means the instance can be solved in less than 0.01 second. If some of the 30 instances cannot be solved within 3 hours, we also append the number of solved instances in brackets. If none of the 30 instances can be solved, for the first three algorithms, we report the average best lower bound, for MIP, we report both the average best lower bound and average upper bound. The shortest times among the first three algorithms and between the two mathematical formulations are highlighted by bold font.
%However, CPLEX may fail to give a result for some large graphs due to the memory limitation.

For the tested random graphs, the time consumption of UBP is insignificant with respect ro the whole search time. Meanwhile, the number of iterations for propagating upper bounds is also trivial (closely around 3 for all the configurations). In terms of computational time of all the algorithms, ExtBBClq is generally the fastest algorithm to solve most of these instances while ExtUniBBClq also performs better than BBClq. The MIP formulation is found to be quite competitive compared with the other three algorithms when the density of the instance reaches 0.9. A possible explanation to this phenomenon is that the formulation of dense instances involves fewer constraints and may be solved more easily. In addition, when graphs density increases much, the maximum biclique and the maximum balanced biclique tend to be closer and closer, which means that the balancing constraint, which makes the problem NP-hard, tends to be less and less active. Since the maximum biclique problem is easy on bipartite graph, a balanced biclique is likely to be found easily by a MIP solver (as the quality of the linear programming relaxation improves with density) whereas high density is the worst situation for enumerative approaches. As expected, the tightened MIP is often solved faster than the original MIP, and the gap to optimality is generally less for those instances that could not be solved to optimality.

%\textcolor{red} {We also find that the tightened formulation accelerates CPLEX for the majority of random graphs, with only two exceptions are observed on small instances rand\_50\_0.3, rand\_50\_0.5 and rand\_100\_0.3.} 

We report the results for the set of large real-life instances in Table \ref{tbl_konect_result}. Column ``instance" indicates the name of graph. Column ``best" shows the best half-size found by all algorithms. An extra ``*" indicates the  optimality of this best value. Column ``UBP" also reports the time of pre-processing and number of iterations to propagate the upper bounds. For each algorithm, we report the computational time to solve the instances. As in Table \ref{tbl_random_result}, when optimality is not proven, the best lower bound is reported for the first three algorithm while for MIP, both lower bound and upper bound are presented.

For the large real-life instances (see Table \ref{tbl_konect_result}), the time spent by UBP is still insignificant with respect to the total search time. The number of iterations is also limited in fewer than 40 for these very large instances. We note that the new algorithms with UBP (ExtBBClq and ExtUniBBClq) dominate the original algorithms in terms of computational time. The extended version ExtBBClq reduces the time of BBClq from hundreds of seconds to less than 30 seconds for 10 instances. It is also the only algorithm that solves discogs\_affiliation. ExtUniBBClq is faster than ExtBBClq on 8 instances. It also achieves a substantial speed-up on jester-1 (where $|U|\ll|V|$). CPLEX is no longer able to give a lower bound for most instances (but we still observe that the tightened formulation leads to a better performance for the 2 solved instances). 
%Thanks to the minimal degree heuristic, the original BBClq finds optimal solution for gottron-reuters in 3 hours while the others fail.

\subsection{The Size of B\&B tree}

In this section, we compare the sizes of B\&B trees generated by the algorithms to solve MBBP instances. Though in BBClq or ExtBBClq, there is no explicit declaration of B\&B nodes, we can treat one call of BBClq() procedure as one enumeration of B\&B node. In CPLEX, the number of nodes in the search tree is directly available. We exclude ExtUniBBClq as the search scheme and branching heuristic are different from BBClq and ExtBBClq.  

\begin{figure}
	\centering\scalebox{0.75}{\includegraphics[scale=0.55]{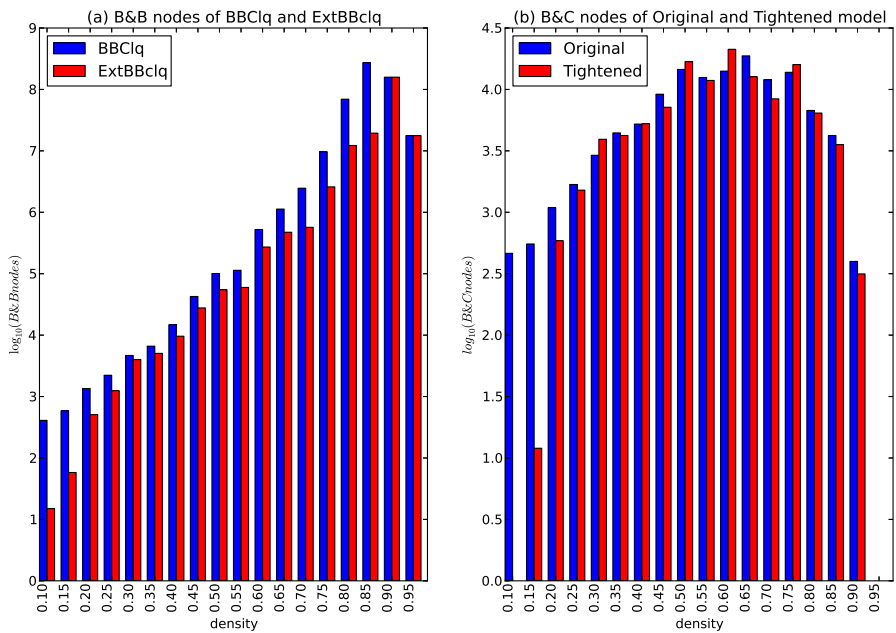}}
	\centering\caption{The base-10 log scale number of B\&B tree nodes explored by BBClq, ExtBBClq, and CPLEX with the original and tightened formulations to solve the random graphs of different densities.} \label{fig_rand_nodes}
	\end{figure}

Firstly, we generate 18 random instances with $n$ fixing to 50 ($|U|=|V|=50$) and $p$ (density) ranging from 0.1 to 0.95, then we solve these instances by BBClq, ExtBBClq and CPLEX with the two formulations. Then we compare the number of B\&B tree nodes between BBClq and ExtBBClq on the one hand, those generated by CPLEX with the original and tightened formulations on the other hand. The results are shown in Figure \ref{fig_rand_nodes}. 

From Figure \ref{fig_rand_nodes}, we can observe that, the bounding technique and the extra inequalities significantly reduce the B\&B tree size on sparse graphs (whose densities are below 0.2). When the density of random graph is lower than 0.9, ExtBBClq always enumerates fewer B\&B nodes than BBClq. However, when the density increase to 0.9, the B\&B tree of ExtBBClq has the same size as that of BBClq. Since we have improved the performance of intersection operation in the implementation of ExtBBClq (see Section \ref{sec_combine_ubp_bbclq}), ExtBBClq still outperforms BBClq when the sizes of B\&B trees are equal (see Table \ref{tbl_random_result}). As to the two mathematical formulations, CPLEX can solve the tightened formulation without expanding the B\&B nodes when the graph is either quite sparse ($p=0.1$) or very dense ($p=0.95$).
	
\begin{figure}
	\centering\scalebox{0.75}{\includegraphics[scale=0.55]{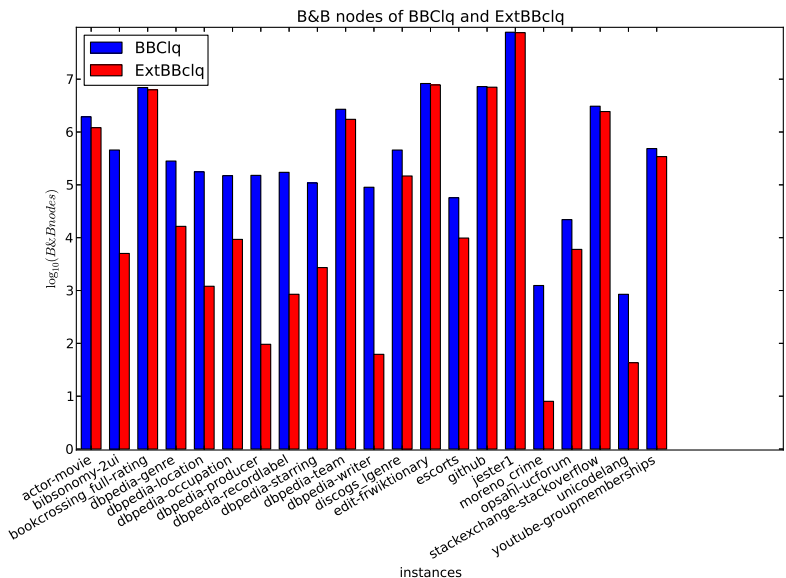}}
	\centering\caption{The base-10 log scaled number of B\&B tree nodes explored by the BBClq and ExtBBClq algorithms for the 21 solvable instances.} \label{fig_real_nodes}
	\end{figure}

Secondly, we compare the sizes of B\&B trees for the real-life instances. Figure \ref{fig_real_nodes} shows the number of B\&B tree nodes of BBClq and ExtBBClq for the 21 instances that can be solved by both algorithms in 3 hours (see Table \ref{tbl_konect_result}). We no longer compare the two mathematical formulations as CPLEX fails to solve the majority of these large instances. Figure \ref{fig_real_nodes} indicates that ExtBBClq enumerates fewer tree nodes for all the real-life instances. This is especially true for  dbpedia-producer, dbpedia-writer and moreno\_crime, where ExtBBClq prunes more than half of the B\&B nodes compared to BBClq.
%but for bookcrossing\_full-rating, edit-frwiktionary, github and jester1 the improvements are not significant.

\section{Conclusion and Future Work}

In this paper, we proposed new ideas for designing exact algorithms for the NP-hard Maximum Balanced Biclique Problem. We introduced the Upper Bound Propagation (UBP) procedure for the sake of estimating tight upper bound involving each vertex. UBP starts from the initial bound of each vertex and improves the upper bounds by a propagating procedure. Based on UBP, we extended the B\&B algorithm (BBClq) of \cite{mccreesh2014exact} and proposed new valid inequalities to tighten the MIP formulation of \cite{dawande2001bipartite}. Furthermore, we presented a new exact algorithm (ExtUniBBClq) which enumerates eligible vertex subsets rather than feasible bicliques like BBClq. Experiments showed the effectiveness of the new proposed ideas for random graphs as well as large real-life instances. Further experiments also confirm that our bounding technique reduces the size of B\&B search tree for the majority of benchmark instances.

As future work, it would be interesting to investigate the bit-parallel technique \cite{san2011exact} within our algorithms to further improve their performances. Also, the branch-and-cut approach based on the tightened formulation constitutes another promising perspective for solving the problem. Finally, the idea of upper bound propagation could be adapted to other similar optimization problems. 

%\bibliography{bbclique}
\bibliographystyle{elsart-num-sort}

\end{document}